\documentclass[prl,nofootinbib,twocolumn,superscriptaddress]{revtex4}
\def\mysection#1{{\bf #1.} }

\usepackage{amssymb}
\usepackage{amsmath}
\usepackage[dvips]{graphicx}
\usepackage{longtable}
\usepackage{verbatim}
\usepackage{amsfonts}

\newcommand{\beq}{\begin{equation}}
\newcommand{\eeq}{\end{equation}}
\newcommand{\beqa}{\begin{eqnarray}}
\newcommand{\eeqa}{\end{eqnarray}}
\newcommand{\no}{\nonumber}

\newcommand{\lsim}{\mathrel{\rlap{\lower4pt\hbox{\hskip1pt$\sim$}}
    \raise1pt\hbox{$<$}}}         %less than or approx. symbol
\newcommand{\gsim}{\mathrel{\rlap{\lower4pt\hbox{\hskip1pt$\sim$}}
    \raise1pt\hbox{$>$}}}         %greater than or approx. symbol
\begin{document}
% \draft

\preprint{{\vbox{\hbox{}\hbox{}\hbox{}
    %\hbox{WIS/10/05-May-DPP}
\hbox{hep-ph/0702151}}}}
\vspace*{0.5cm}
\title{Relating leptogenesis parameters to light neutrino masses}

\author{Guy Engelhard}\email{guy.engelhard@weizmann.ac.il}
\affiliation{Department of Particle Physics,
  Weizmann Institute of Science, Rehovot 76100, Israel}
\author{Yuval Grossman}\email{yuvalg@physics.technion.ac.il}
\affiliation{Department of Physics, Technion-Israel Institute of
  Technology, Technion City, Haifa 32000, Israel}
\author{Yosef Nir}\email{yosef.nir@weizmann.ac.il}
\affiliation{Department of Particle Physics,
  Weizmann Institute of Science, Rehovot 76100, Israel}
\date{\today}
%\pacs{}
% \vspace{2cm}

\begin{abstract}
  We obtain model independent relations among neutrino masses and
  leptogenesis parameters. We find exact relations that involve the CP
  asymmetries $\epsilon_{N_\alpha}$, the washout parameters $\tilde
  m_\alpha$ and $\theta_{\alpha\beta}$, and the neutrino masses $m_i$
  and $M_\alpha$, as well as powerful inequalities that involve just
  $\tilde m_\alpha$ and $m_i$.  We prove that the Yukawa interactions
  of at least two of the heavy singlet neutrinos are in the strong
  washout region ($\tilde m_\alpha\gg10^{-3}\ eV$).
\end{abstract}

\maketitle

%%%%%%%%%%%%%%%%%%%%%%%%%%%%
\mysection{Introduction}
\label{sec:introduction}
Singlet neutrinos with heavy Majorana masses and with Yukawa couplings
to the active neutrinos generate light neutrino masses via the see-saw
mechanism and a baryon asymmetry via leptogenesis
\cite{Fukugita:1986hr}, providing attractive qualitative solutions to
these two important puzzles. To be quantitatively successful, the
see-saw mechanism should lead to the two observed mass scales,
\beqa\label{msma}
m_s&\equiv&(\Delta m^2_{\rm sol})^{1/2}\sim0.009\ eV,\no\\
m_a&\equiv&(\Delta m^2_{\rm atm})^{1/2}\sim0.05\ eV,
\eeqa
while leptogenesis should lead to the value extracted from observations,
\begin{equation}
  \label{eq:4}
 Y_{\mathcal{B}}^{\rm obs}\equiv \frac{n_B-n_{\overline{B}}}{s}
 = (8.7\pm 0.3)\times 10^{-11}.
\end{equation}
Unfortunately, because the leptogenesis parameters -- the CP
asymmetries and the washout factors -- directly involve the heavy
singlet neutrinos, we cannot realistically hope that they will be
measured. In order to make further progress in the investigation of
leptogenesis, it is highly desirable to relate the leptogenesis
parameters to measurable mass parameters. The purpose of this work is
to obtain such relations.

The relations that we obtain involve the washout parameters of all the
heavy singlet neutrinos $N_\alpha$. While most leptogenesis studies
have focussed on the contributions from the decays of $N_1$, the
lightest heavy singlet, it has been realized that, in general, the
contributions from the decays of the heavier singlet neutrinos must
not be neglected 
\cite{Barbieri:1999ma,Strumia:2006qk,Engelhard:2006yg}. Indeed, our
results reinforce this statement.

%%%%%%%%%%%%%%%%%%%%%%%%%%%%
\mysection{Notations}
\label{sec:notations}
The relevant Lagrangian terms involve, in addition to the
$N_\alpha$'s, the light lepton SU(2)-doublets $L_i$ and
SU(2)-singlets $E_i$ ($i=e,\mu,\tau$ is a flavor index), and the
standard model Higgs $H$: 
\begin{equation}\label{eq:1}
-\mathcal{L} =\frac{1}{2} M_\alpha N_\alpha N_\alpha + 
\lambda_{\alpha i}HN_\alpha L_i + Y_i
 H^\dagger L_i E_i.
\end{equation}
Eq.~(\ref{eq:1}) is written in the mass basis for the singlet
neutrinos and for the charged leptons, that is, $M$ and $Y$ are
diagonal.

The light neutrino mass matrix is given by
\beq\label{lmnu}
m_\nu=v^2 \lambda^T M^{-1} \lambda,
\eeq
where $v=\langle H\rangle$. Reversing this relation, one can express
the Yukawa couplings $\lambda_{\alpha i}$ in terms of the diagonal
mass matrix $M$, the matrix $m={\rm diag}(m_1,m_2,m_3)$ (where $m_i^2$
are the eigenvalues of $m_\nu m_\nu^\dagger$), the leptonic mixing
matrix $U$ and an orthogonal complex matrix $R$ \cite{Casas:2001sr}:
\beq\label{caib}
\lambda=\frac{1}{v} M^{1/2}\ R\ m^{1/2}\ U^\dagger.
\eeq

The baryon number generated from the decays of the $N_{\alpha}$
neutrinos can be written as follows:
\begin{equation}
  \label{eq:2}
  Y_{\mathcal{B}}= -1.4\times 10^{-3}\sum_{\alpha,\beta}
\epsilon_{N_\alpha}\eta_{\alpha\beta},
\end{equation}
where $\epsilon_{N_\alpha}$ is the CP asymmetry generated in
$N_\alpha$ decays: 
\begin{equation}\label{eq:eps}
\epsilon_{N_\alpha}=\frac{\Gamma(N_\alpha\to\ell H)-\Gamma(N_\alpha\to\bar\ell
  \bar H)}{\Gamma(N_\alpha\to\ell H)+\Gamma(N_\alpha\to\bar\ell \bar H)},
\end{equation}
and $\eta_{\alpha\beta}$ denotes the efficiency factor related to the
washout of the asymmetry $\epsilon_{N_\alpha}$ due to $N_\beta$
interactions. (If leptogenesis takes place at $T\lsim10^{12}$, flavor
indices should be added
\cite{Barbieri:1999ma,Endoh:2003mz,Nardi:2006fx,Abada:2006fw}.)  It is convenient
for our purposes to further define a matrix of dimensionful quantities
$\tilde m_{\alpha\beta}$:
\begin{equation}
  \label{eq:tmi}
  \tilde m = v^2 M^{-1/2}\lambda\lambda^{\dag} M^{-1/2}. 
\end{equation}
Note that $\tilde m$ is a positive matrix and, in particular, $|\tilde
m_{\alpha\beta}|^2\leq\tilde m_{\alpha\alpha}\tilde m_{\beta\beta}$.
In terms of the parametrization (\ref{caib}), we have
\beq\label{tmcaib}
\tilde m_{\alpha\beta}=\sum_i m_i R_{\alpha i}R_{\beta i}^*.
\eeq
In a large part of the parameter space, the washout factors
$\eta_{\alpha\alpha}$ depend on the mass and the couplings of
$N_\alpha$ only via the combination $\tilde m_\alpha\equiv \tilde
m_{\alpha\alpha}$ \cite{Buchmuller:2002rq}. For example, for
$M_1\ll10^{14}\ GeV$ and $\tilde m_\alpha\gg m_*=2.2\times10^{-3}\ 
eV$, we have \cite{Giudice:2003jh}
\beq\label{etatm}
\eta_{\alpha\alpha}\approx \left(\frac{5.5\times10^{-4}\ eV}{\tilde
    m_\alpha}\right)^{1.16}.
\eeq

When we talk in this work about the ``washout parameters'' we refer
mainly to the $\tilde m_\alpha$'s. The off-diagonal terms in
$\tilde m$ do, however, play important roles in leptogenesis. First,
the CP asymmetries depend on ${\cal I}m(\tilde m_{\alpha\beta})$ (see,
for example, Eq. (\ref{epsim}) below). Second, $|\tilde
m_{\alpha\beta}|$ determines the overlap between the lepton doublet
states $\ell_\alpha$ and $\ell_\beta$ to which $N_\alpha$ and
$N_\beta$ decay, respectively \cite{Engelhard:2006yg}:
\beqa
|\ell_\alpha\rangle&=&(\lambda\lambda^\dagger)_{\alpha\alpha}^{-1/2}
\sum_i \lambda_{\alpha i}|\ell_i\rangle,\no\\
\cos^2\theta_{\alpha\beta}&\equiv&|\langle\ell_\alpha|\ell_\beta\rangle|^2=
|\tilde m_{\alpha\beta}|^2/(\tilde m_\alpha\tilde m_\beta).
\eeqa
For the case of strong hierarchy between the masses and the lifetimes
of, say, $N_1$ and $N_2$, and $\tilde m_1\gg m_*$, the interactions of
$N_1$ first project $\epsilon_{N_2}$ on the directions aligned with
or orthogonal to $\ell_1$ and then washout the asymmetry in the
$\ell_1$ direction \cite{Engelhard:2006yg}. For this case, we
use an approximate expression for the total lepton asymmetry
generated in $N_2$ and $N_1$ decays:
\beq\label{epsonetwo}
Y_{\mathcal{B}}\approx-1.4\times10^{-3}\left[\epsilon_{N_1}\eta_{11}
  +\epsilon_{N_2}\eta_{22}(\cos^2\theta_{12}\eta_{11}
    +\sin^2\theta_{12})\right].
\eeq

%%%%%%%%%%%%%%%%%%%%%%%%%%%%%%%%%%%%%%%%%%%%%%%%%%%%
\mysection{The basic relations}
The key point for our results is that $\tilde m^*\tilde m$ and $m_\nu
m_\nu^\dagger$ are similar. In particular, the following three
relations hold:
\beqa\label{mnutm}
\det(m_\nu m_\nu^\dagger)&=&\det(\tilde m^*\tilde m),\no\\
{\rm Sym}_2(m_\nu m_\nu^\dagger)&=&{\rm Sym}_2(\tilde m^*\tilde m),\no\\
{\rm Tr}(m_\nu m_\nu^\dagger)&=&{\rm Tr}(\tilde m^*\tilde m),
\eeqa
where ${\rm Sym}_2(A)=\frac12\left\{\left[{\rm Tr}(A)\right]^2-{\rm
    Tr}(A^2)\right\}$.

These equations can be written as exact relations involving the light
neutrino masses $m_i$, the washout parameters $\tilde m_\alpha$, and
the off-diagonal terms $\tilde m_{\alpha\beta}$. The latter can be
expressed in terms of the CP asymmetries $\epsilon_{N_\alpha}$ and the
projections $\cos^2\theta_{\alpha\beta}$.

These equalities [as well as the explicit form (\ref{tmcaib})] can be
further used to obtain simple inequalities involving only the washout
parameters $\tilde m_\alpha$ and the light neutrino masses $m_i$. In
particular, we are able to show that some (and in some cases all) of
the $N_\alpha$ interactions are in the strong washout region.

We note that there is no additional information for us in the leptonic 
mixing angles. The reason is that $\tilde m$ is independent of the
mixing angles. This can be seen by noting that
$\lambda\lambda^\dagger$ is independent of $U$, Eq. (\ref{caib}), or
directly from Eq. (\ref{tmcaib}). 

We apply Eqs. (\ref{mnutm}) to two cases, differing in the number of
singlet neutrinos $N_\alpha$ that are added to the SM. In the ``3+2''
framework, two such neutrinos are assumed to be relevant to the
see-saw mechanism and to leptogenesis, while in the ``3+3'' framework,
there are three. The $3+2$ case is actually a special limit of the
$3+3$ framework. When one of the three $\tilde m_\alpha\to0$ (that is,
$(\lambda\lambda^\dagger)_{\alpha\alpha}\to0$ and/or
$M_\alpha\to\infty$), the corresponding $N_\alpha$ becomes irrelevant
to both the see-saw mechanism and leptogenesis, and the model reduces
to effectively a $3+2$ model. We checked that all our results in the
$3+3$ framework indeed reproduce the $3+2$ results in this limit.

We do not consider $3+m$ models with $m>3$ because, in general, they
are similar to the $3+3$ model. To understand that, note that
$\ell_2$ ($\ell_3$) has, in general, a component that is orthogonal to
$\ell_1$ ($\ell_1$ and $\ell_2$). Consequently, part of the
asymmetries generated by the decays of $N_2$ and $N_3$ is protected
against washout \cite{Engelhard:2006yg}. The light flavor space is,
however, three dimensional and therefore spanned, in general, by
$\ell_1$, $\ell_2$ and $\ell_3$. Consequently, there is no component
in $\ell_{\alpha>3}$ that is orthogonal to all three, and the
asymmetries $\epsilon_{N_{\alpha>3}}$ are expected to be washed out. 

%%%%%%%%%%%%%%%%%%%%%%%%%%%%%%%%%%%%%%%%%%%%%%%
\mysection{The $3+2$ framework}
When we have only two singlet neutrinos [$\alpha=1,2$ in Eq.
(\ref{eq:1})], one of the light mass eigenvalues vanishes and the
other two mass eigenvalues are fixed by phenomenology to one of two
discrete possibilities, either normal hierarchy (NH) or inverted
hierarchy (IH):
\beqa\label{masstt}
m_1&=&0,\ \ m_2=m_s,\ \ m_3=m_a\ \ ({\rm NH}),\\
m_1&=&0,\ \ m_{2,3}\approx m_a,\ \ m_3-m_2=m_s^2/(2m_a)\ \
({\rm IH}).\no
\eeqa

Using Eqs. (\ref{mnutm}), we have
\beqa\label{prodmi}
m_2m_3&=&\tilde m_1\tilde m_2-|\tilde m_{12}|^2,\\
\label{summi}
m_2^2+m_3^2&=&\tilde m_1^2+\tilde m_2^2+2{\cal R}e(\tilde m_{12}^2).
\eeqa
Using Eqs. (\ref{prodmi}) and (\ref{summi}), we obtain two inequalities:
\beqa\label{sumin}
\tilde m_2+\tilde m_1&\geq& m_3+m_2,\\
|\tilde m_2-\tilde m_1|&\leq& m_3-m_2.\label{difin}
\eeqa
We derived various additional inequalities:
\beqa\label{lowtm}
\tilde m_{1,2}&\geq& m_2,\\
\label{proin}
\tilde m_1\tilde m_2&\geq& m_2m_3,\\
\label{ratlim}
m_2/m_3\leq\tilde m_1/\tilde m_2&\leq& m_3/m_2.
\eeqa
(The inequality $\tilde m_1\geq m_2$ was derived in
ref. \cite{Ibarra:2003up}.) 
However, Eqs. (\ref{lowtm})-(\ref{ratlim}) are redundant when the
constraints (\ref{sumin}) and (\ref{difin}) are imposed.

\begin{figure}[bt]
  \centering
  \includegraphics[scale=0.8]{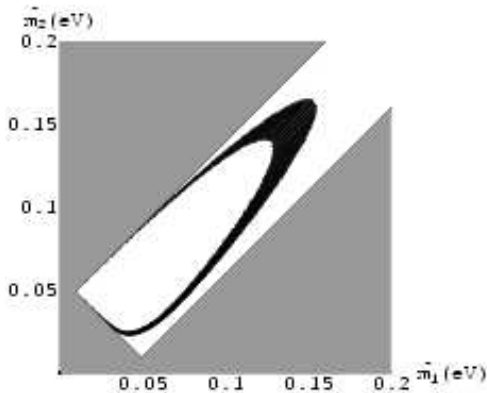}
  \caption{The constraints in the $\tilde m_1-\tilde m_2$ plane in the
    $3+2$ case with normal hierarchy. The grey region is forbidden by
    Eqs. (\ref{sumin}) and (\ref{difin}). The black region is derived
    by scanning the parameter space (fixing $M_1=10^{12}\ GeV$ and
    $M_2/M_1=10$) and requiring that the resulting baryon asymmetry
    would be within the $3\sigma$ range [Eq. (\ref{eq:4})].}
  \label{fig:momt}
\end{figure}

The combination of the inequalities (\ref{sumin}) and (\ref{difin}),
together with the known values of $m_2$ and $m_3$ [Eq.
(\ref{masstt})], constrains the allowed region in the $\tilde
m_1-\tilde m_2$ plane in a very significant way. We plot these
constraints in Fig. \ref{fig:momt}. In particular, we can draw the following
conclusions:
\begin{enumerate}
\item For NH, both $\tilde m_\alpha$ are above $m_s$ and at
  least one of them is above $m_a/2$. The two washout factors are
  within a factor of $m_a/m_s\sim6$ of each other.
\item For IH, both $\tilde m_\alpha$ are above $m_a$, while
  the difference between them is very small, $\leq m_s^2/(2m_a)$.
\item In either case, both $N_2$ interactions and $N_1$ interactions
  are in the strong washout regime.
      \end{enumerate}

As concerns the two CP asymmetries, $\epsilon_{N_{1,2}}$, they can 
be written as ($x_{12}\equiv M_1/M_2$)
\beq\label{epsim}
\epsilon_{N_{\alpha}}=f_\alpha(x_{12})\frac{M_\alpha{\cal I}m(\tilde m_{12}^2)}
{v^2\tilde m_{\alpha}}.
\eeq
The functions $f_\alpha(x_{12})$ can be found in the literature
\cite{Covi:1996wh}.
Instead of the inequalities (\ref{sumin}) and (\ref{difin}),
one can combine Eqs. (\ref{prodmi}), (\ref{summi}) and (\ref{epsim})
to obtain an {\it exact} relation between the neutrino
masses, the washout parameters and the CP asymmetries:
\beqa\label{exarel}
4(\tilde m_1\tilde m_2-m_2m_3)^2&-&(\tilde m_1^2+\tilde
m_2^2-m_2^2-m_3)^2\no\\
=4\frac{v^4\tilde m_1^2\epsilon_{N_1}^2}{M_1^2[f_1(x_{12})]^2}
&=&4\frac{v^4\tilde m_2^2\epsilon_{N_2}^2}{M_2^2[f_2(x_{12})]^2}.
\eeqa

%%%%%%%%%%%%%%%%%%%%%%%%%%%%%%%%%%%%%%%%%%%%%%%
\mysection{The $3+2$ framework with strong $M_2/M_1$ hierarchy}
While our main focus here is on model independent relations, we can
gain some further understanding by assuming mass hierarchy between the
two singlet neutrinos. (It also explains features of the black region
in Fig. \ref{fig:momt} which corresponds to $M_2/M_1=10$.) 
In the hierarchical case ($x_{12}\ll1$), we have
\beqa
f_1(x_{12})&=&-3/(16\pi),\no\\
f_2(x_{12})&=&-x_{12}^2[\ln(x_{12})+1]/(4\pi),\no\\
\frac{\epsilon_{N_2}}{\epsilon_{N_1}}&=&-\frac43\frac{\tilde m_1}{\tilde
  m_2}x_{12}\left(\ln\frac{1}{x_{12}}-1\right).
\eeqa
Using Eqs. (\ref{eq:2}) and (\ref{etatm}), we can give a rough
estimate of the ratio between the respective contributions to $Y_{\mathcal{B}}$:
\beq
\frac{|\epsilon_{N_2}|/\tilde m_2}{|\epsilon_{N_1}|/\tilde m_1}\sim\frac{\tilde m_1^2}{\tilde
  m_2^2}\frac{M_1}{M_2}\left(\ln\frac{M_2}{M_1}-1\right).
\eeq
We would like to emphasize the following points:
\begin{enumerate}
\item For a mild hierarchy between $M_1$ and $M_2$, $N_2$-leptogenesis
  must not be neglected. (In this case, $\eta_{12}$ and $\eta_{21}$
  have to be taken into account.)
  \item In the NH case, only for a very strong hierarchy,
    $M_2/M_1\gg10^2$, it is guaranteed that $N_1$ leptogenesis
    dominates. For IH, the contribution from $\epsilon_{N_1}$ is
    always larger.
  \item Given that $\epsilon_{N_2}$ and $\epsilon_{N_1}$ have opposite
    signs, partial (and potentially significant) cancellation between
    the two contributions is quite possible.
      \end{enumerate}

In the $3+3$ framework, when the heavy neutrino masses are strongly
hierarchical, $\epsilon_{N_1}$ is subject to an upper bound
\cite{Davidson:2002qv,Hambye:2003rt}:
\beq\label{dibou}
|\epsilon_{N_1}|\leq\epsilon^{\rm DI}\equiv
\frac{3}{16\pi}\frac{M_1 (m_3-m_2)}{v^2}.
\eeq
We first note that an even stronger bound applies to
$|\epsilon_{N_2}|$: 
\beq
|\epsilon_{N_2}|\leq\frac{4}{3}\frac{M_1}{M_2}
\left[\ln\left(\frac{M_2}{M_1}\right)-1\right]\epsilon^{\rm DI}.
\eeq
Second, we note that $m_3-m_2$ is fixed to either $m_a-m_s$ (NH) or
$m_s^2/(2m_a)$ (IH) \cite{Chankowski:2003rr}, and so we can be more
specific in Eq. (\ref{dibou}): 
\beq
\epsilon^{\rm DI}=\begin{cases}
  M_1/(2.5\times10^{16}\ GeV)&{\rm NH}\cr
  M_1/(10^{18}\ GeV)&{\rm IH}\cr \end{cases}  
\eeq
In addition, since for NH(IH) $\tilde m_1\geq m_s(m_a)$, we have
$\eta_{11}\leq0.04(0.006)$. The upper bounds on $\epsilon_{N_1}$ and on
$\eta_{11}$ give lower bounds on $M_1$,
\beq
M_1\gsim\begin{cases}
  3.6\times10^{10}\ GeV&{\rm NH}\cr
  1.3\times10^{13}\ GeV&{\rm IH}\cr
  \end{cases}
\eeq
Alternatively, one can write upper bounds on $\tilde m_1$:
\beq
\tilde m_1\lsim\begin{cases}
  m_s\times[M_1/(3.6\times10^{10}\ GeV)]^{0.86}&{\rm NH}\cr
  m_a\times[M_1/(1.3\times10^{13}\ GeV)]^{0.86}&{\rm IH}\cr
\end{cases}
\eeq

%%%%%%%%%%%%%%%%%%%%%%%%%%%%%%%%%%%%%%%%%%%%%%%
\mysection{The $3+3$ framework}
Within the $3+3$ framework, we can distinguish three different types
of light neutrino spectra, Normal hierarchy (NH), inverted hierarchy
(IH), and quasi degeneracy (QD): 
\beqa
m_1&\ll& m_s,\ \ m_2\approx m_s,\ \ m_3\approx m_a\ \ ({\rm NH});\no\\
m_1&\ll& m_a,\ \ m_{2,3}\approx m_a,\ \ m_3-m_2=\frac{m_s^2}{2m_a}\ \ ({\rm IH});\no\\
m_i&\approx&\bar m\gg m_a,\ \ m_3-m_2=\frac{m_a^2}{2\bar m},\ \
m_2-m_1=\frac{m_s^2}{2\bar m}\ \ ({\rm QD}).\no
\eeqa

Using Eq. (\ref{tmcaib}), we obtain lower bounds on the washout
parameters (the first relation was derived in
Ref. \cite{Fujii:2002jw}):  
\beqa\label{sintm}
\tilde m_\alpha&\geq& m_1,\\
\label{trtmit}
\tilde m_1+\tilde m_2+\tilde m_3&\geq& m_1+m_2+m_3.
\eeqa
Evaluating Eqs. (\ref{mnutm}), we obtain
\beq\label{trthree}
m_1^2+m_2^2+m_3^2
=\tilde m_{1}^2+\tilde m_{2}^2+\tilde m_{3}^2+2{\cal R}e(\tilde
m_{12}^2+\tilde m_{23}^2+\tilde m_{13}^2),
\eeq
\beqa\label{symtwo}
m_1^2m_2^2&+&m_1^2m_3^2+m_2^2m_3^2
=(\tilde m_{11}\tilde m_{22}-|\tilde m_{12}|^2)^2\no\\
&+&(\tilde m_{11}\tilde m_{33}-|\tilde m_{13}|^2)^2
+(\tilde m_{22}\tilde m_{33}-|\tilde m_{23}|^2)^2\no\\
&+&2{\cal R}e\left[(\tilde m_{11}\tilde m_{23}^*-\tilde m_{12}\tilde m_{31})^2\right.\\
&+&\left.(\tilde m_{22}\tilde m_{31}^*-\tilde m_{23}\tilde m_{12})^2
+(\tilde m_{33}\tilde m_{12}^*-\tilde m_{31}\tilde m_{23})^2\right].\no
\eeqa
\beqa\label{dettrtr}
m_1m_2m_3
&=&\tilde m_{11}\tilde m_{22}\tilde m_{33}-\tilde m_{33}|\tilde m_{12}|^2
-\tilde m_{22}|\tilde m_{13}|^2\no\\
&-&\tilde m_{11}|\tilde m_{23}|^2
+2{\cal R}e(\tilde m_{12}\tilde m_{23}\tilde m_{31})\no\\
&\leq&\tilde m_1\tilde m_2\tilde m_3.
\eeqa
Using the fact that $\tilde m$ is hermitian and positive, and the
general property that for any positive definite matrix $A$ one has 
Tr$[(AA^\dagger)^{-1/2}]\geq{\rm Tr}[(AA^*)^{-1/2}]$ \cite{Oded}, we obtain 
\beqa\label{yuvcon}
\tilde m_1\tilde m_2+\tilde m_2\tilde m_3&+&\tilde m_3\tilde
m_1-(m_1m_2+m_2m_3+m_3m_1)\no\\
&\geq&
|\tilde m_{12}|^2+|\tilde m_{23}|^2+|\tilde
m_{13}|^2.
\eeqa
Eqs. (\ref{trthree}) and (\ref{yuvcon}) can be combined to give
\beqa\label{yuvine}
\tilde m_1^2&+&\tilde m_2^2+\tilde m_3^2
-2(\tilde m_1\tilde m_2 + \tilde m_1\tilde m_3 + \tilde m_2\tilde
m_3)\\
&\leq&m_1^2+m_2^2+m_3^2-2(m_1m_2 + m_1m_3 + m_2m_3).\no
\eeqa

We now use the above equations and inequalities to obtain lower bounds
on the $\tilde m_\alpha$'s. We denote by $\tilde m_a,\tilde m_b,\tilde
m_c$ the smallest, intermediate and largest $\tilde m_\alpha$,
respectively. The left hand side of the inequality
(\ref{yuvine}) is minimized for $\tilde m_b-\tilde m_a=0$. It is
maximized (for $\tilde m_c\geq 2(\tilde m_b+\tilde m_a)$) by maximal
$\tilde m_c$ which, according to Eq. (\ref{trtmit}), is given by $\sum_i
m_i-(\tilde m_b+\tilde m_a)$. We can then write an inequality that
depends on the sum of the two smallest $\tilde m_\alpha$:
\beq
3(\tilde m_a+\tilde m_b)^2-4\sum_i m_i(\tilde m_a+\tilde
m_b)+4\sum_{i<j}m_im_j\leq0,
\eeq
This inequality leads to an interesting lower bound,
\beq\label{lowsum}
\tilde m_a+\tilde m_b\geq \frac23\sum_i m_i-\frac23\left(\sum_i
m_i^2-\sum_{i<j}m_im_j\right)^{1/2}.
\eeq
This bound has interesting implications for models of $N_2$
leptogenesis that are based on $\tilde m_1$ in the weak washout region
\cite{Vives:2005ra,DiBari:2005st,Blanchet:2006dq}, where it gives
$\tilde m_2\gsim m_s$. (A qualitative statement in this regard was
made in ref. \cite{DiBari:2005st}.)

Eqs. (\ref{sintm}), (\ref{trtmit}) and (\ref{lowsum}) lead to the
following lower bounds:
\begin{itemize}
\item Normal hierarchy:
\beq
\tilde m_c\geq\frac{m_a}{3}\left(1+\frac{m_s}{m_a}\right),\ \
\tilde m_b\geq\frac{m_s}{2}\left(1-\frac{m_s}{4m_a}\right),\ \ 
\tilde m_a\geq m_1.
\eeq
\item Inverted hierarchy:
\beq
\tilde m_c\geq\frac{2m_a}{3},\ \  
\tilde m_b\geq\frac{m_a}{2},\ \
\tilde m_a\geq m_1.
\eeq
\item Quasi degeneracy:
\beq
\tilde m_\alpha\geq\bar m.
\eeq
\end{itemize}

We conclude that, for hierarchical (quasi-degenerate) light neutrino
masses, at least two (all three) of the $\tilde m_\alpha$ are in the strong
washout region.

%%%%%%%%%%%%%%%%%%%%%%%%%%%%%%%%%%%%%%%%%%%%%%%%
\mysection{Conclusions} 
\label{sec:conclusions} 
We investigated the relations between leptogenesis parameters and
light neutrino masses. In particular, we derived exact relations
between elements of the $\tilde m$ matrix [defined in Eq.
(\ref{eq:tmi})], relevant to leptogenesis, and the light neutrino
masses.  The diagonal elements, $\tilde m_{\alpha}$, determine the
$\Delta L=1$ washout effects. As concerns the off-diagonal ones,
${\cal I}m(\tilde m_{\alpha\beta})$ determine the size of the CP
asymmetries, while $|\tilde m_{\alpha\beta}|$ is related to
projections (in heavy flavor space) of the asymmetries generated by
heavy singlet neutrinos due to interactions of lighter singlets.

The resulting equations lead to interesting exact relations, such as
Eq. (\ref{exarel}), between the washout parameters, CP asymmetries and
neutrino masses. The various relations lead to simple inequalities
between the washout parameters $\tilde m_\alpha$ and the light neutrino
masses $m_i$, see Eqs. (\ref{sumin})--(\ref{ratlim}) for the $3+2$
framework and (\ref{sintm})--(\ref{lowsum}) for the $3+3$ framework.

For light neutrino masses with normal hierarchy, we find the following
results:
\begin{itemize}
\item In the $3+2$ framework, both $N_1$ and $N_2$ interactions are in
  the strong washout region, with both $\tilde m_\alpha\geq 0.009$ eV and
  at least one $\geq0.025$ eV.
\item In the $3+3$ framework, at least two $N_\alpha$'s have
  interactions in the strong washout region, with $\tilde m_\alpha\geq
  0.005$ eV and at least one $\geq0.02$ eV.
\end{itemize}
The lower bounds are stronger for inverted hierarchy, and even more so
in the $3+3$ framework with quasi-degenerate light neutrinos.

\smallskip\noindent
%%%%%%%%%%%%%%%%%%%%%%%
{\bf Acknowledgments.}
We thank Micha Berkooz, Oded Kenneth, Enrico Nardi, Esteban Roulet,
Adam Schwimmer and Alessandro Strumia for useful discussions. This
project was supported by the Albert Einstein Minerva Center for
Theoretical Physics. The work of Y.G. is supported in part by the
Israel Science Foundation under Grant No. 378/05. The research of Y.N.
is supported by grants from the Israel Science Foundation founded by the
Israel Academy of Sciences and Humanities, the United States-Israel
Binational Science Foundation (BSF), Jerusalem, Israel, the
German-Israeli foundation for scientific research and development
(GIF), and the Minerva Foundation.

%%%%%%%%%%%%%%%%%%%%%%%%%%%%%

\end{document}